\documentclass[11pt,a4paper]{article}

\usepackage[affil-it,blocks]{authblk}
\usepackage[utf8]{inputenc}
\usepackage{amsmath}

\def\tref{h_{\rm (r)}}
\def\Lbol{{\stackrel{\circ}{\mathcal L}}{}}

\def\D{{{\mathcal D}}{}}
\def\Lw{{\mathcal L}{}}
\def\ombol{{\stackrel{\circ}{\omega}}{}}

\usepackage[numbers,compress]{natbib}

\begin{document}
\title{ \bf  Variational Problem and Bigravity Nature of Modified Teleparallel Theories }
\author{ Martin Kr\v{s}\v{s}\'ak\thanks{Electronic address: \texttt{martin.krssak@ut.ee}} }

\affil{Institute of Physics, University of Tartu,\\ W. Ostwaldi 1, Tartu 50411, Estonia}

\maketitle
\begin{abstract}
We  consider the  variational principle in  the covariant formulation of modified teleparallel theories with second order field equations. We vary the action with respect to the spin connection and  obtain a consistency condition relating the spin connection with the tetrad.  We argue that since the spin connection can be calculated using  an additional reference tetrad, modified  teleparallel theories can be interpreted as effectively  bigravity theories. We conclude with discussion about the relation of our results and those obtained in the usual, non-covariant, formulation of teleparallel theories and present the solution to the problem of choosing the tetrad in $f(T)$ gravity theories. 
\end{abstract}

\vskip 0.8cm
\section{Introduction}
Teleparallel gravity is an alternative formulation of general relativity that can be traced back to Einstein's attempt to formulate the unified field theory \cite{Einstein1928,Sauer:2004hj,Moller1961,Hayashi:1967se,Cho:1975dh,Hayashi:1977jd, Maluf:1994ji,AP,Maluf:2013gaa}.  Over the last decade, various modifications of teleparallel gravity became a popular tool to address  the problem of the  accelerated expansion of the Universe without invoking the dark sector  \cite{Ferraro:2006jd,Ferraro:2008ey,Bengochea:2008gz,Linder:2010py,Wu:2010mn,Bengochea:2010sg,Dent:2011zz,Bamba:2010wb,Capozziello:2011hj,Wu:2011kh,Farajollahi:2011af,Cardone:2012xq,Bamba:2012vg,Nesseris:2013jea,Harko:2014sja,Geng:2014nfa,Bamba:2014eea,Hanafy:2014ica,Capozziello:2015rda,Nashed:2015pda,Bahamonde:2015zma,Jarv:2015odu,Farrugia:2016xcw,Oikonomou:2016jjh,Wu:2016dkt,Farrugia:2016qqe,Nunes:2016qyp,Nunes:2016plz, Capozziello:2017bxm}.    The attractiveness of modifying  teleparallel gravity--rather than the usual general relativity--lies in the fact that we obtain an entirely new class of modified gravity theories, which have  second order field equations. See \cite{Cai:2015emx} for the extensive review.

A well-known shortcoming of the original formulation of teleparallel theories is the problem of local Lorentz symmetry violation \cite{Li:2010cg,Sotiriou:2010mv}. This is particularly serious in the modified case,  where we often encounter situations that out of  two tetrads related by a local Lorentz transformation, only one solves the field equations. Since  both tetrads correspond to the same metric, it is  not the metric tensor but a specific tetrad that solves a problem in modified teleparallel theories.

This introduces  a new complication of how to choose the tetrad.
As it turns out, the tetrad must be chosen in a specific way or otherwise the field equations lead us to the condition $f_{TT}=0$,  which restricts the  theory to its general relativity limit. Tetrads that lead to this restriction were nicknamed \textit{bad tetrads}, while those that avoid it and lead to some interesting new dynamics as  \textit{good tetrads} \cite{Ferraro:2011us,Tamanini:2012hg}.   An intriguing sub-class of good tetrads are those that lead to $T=\mathrm{const}$,  which can be used to show that the solutions of the ordinary GR are the universal solutions of an arbitrary $f(T)$ gravity \cite{Ferraro:2011ks,Bejarano:2014bca}.

Recently \cite{Krssak:2015oua}, it was shown that local Lorentz invariance can be restored in  modified teleparallel theories, if the starting point is  a more general formulation of teleparallel gravity  \cite{Obukhov:2002tm,Obukhov:2006sk,Lucas:2009nq,Blagojevic,Hehl:1994ue,Krssak:2015rqa,Krssak:2015lba,Golovnev:2017dox}, in which  teleparallelism is defined by a condition of vanishing curvature.  This introduces the purely inertial spin connection to the theory that can be calculated consistently  in the  ordinary teleparallel gravity  \cite{Krssak:2015rqa,Krssak:2015lba}.  However, in the modified case it was argued to be possible only to ``guess" the connection by making  some reasonable assumptions about the asymptotic properties of the ansatz tetrad  \cite{Krssak:2015oua}. While it was shown to work, and even to be rather easily achievable, in many physically interesting cases \cite{Krssak:2015oua,DeBenedictis:2016aze,Otalora:2017qqc,Velay-Vitow:2017odc},  it is obviously   not a satisfactory situation and better understanding  of covariant modified teleparallel theories is needed. 

In this paper, we discuss various approaches to the  the variational principle and derive the condition (\ref{concon}) relating  the spin connection with tetrad by varying the action with respect to the spin connection. We argue that in many situations the solutions of this condition correspond to the ``guesses"  discussed in Ref.~\cite{Krssak:2015oua}. We further obtain  understanding of why  $T=\text{const}$ can be used to show the universality of general relativity solutions in $f(T)$ gravity. We show that all these results can be always interpreted in the original, non-covariant, formulation and explain the physical origin of why some tetrads are ``good", and some other ``bad".

Moreover, we argue that we can view modified teleparallel gravity theories as effectively bigravity theories with two tetrads, where the first tetrad determines the spacetime metric, while  the second tetrad generates the spin connection and corresponds to the non-dynamical, ``reference", metric.

\textit{Notation:} We follow notation, where the Latin indices $a,b,...$ run over the tangent space, while the Greek indices $\mu,\nu,...$ run over spacetime coordinates. The spacetime indices are raised/lowered using the spacetime metric $g_{\mu\nu}$, while the tangent indices using the Minkowski metric $\eta_{ab}$ of the tangent space.  
 
\section{Covariant formulation of teleparallel theories} 
Teleparallel theories are formulated in the framework of  tetrad formalism, where the fundamental variable is  the tetrad, $h^a_{\ \mu}$, related to the spacetime  metric  through the relation
\begin{equation}
g_{\mu\nu}=\eta_{ab}h^a_{\ \mu} h^b_{\ \nu}.\label{met}
\end{equation}
The parallel transport is determined by the spin connection, which is fully  characterized by its  curvature and torsion in the metric compatible case. 
In general relativity, we use the connection with vanishing  torsion known as the Levi-Civita connection. In teleparallel gravity, we follow a complimentary approach and define the  teleparallel connection, $\omega^a_{\ b\mu}$, by the condition of zero curvature
\begin{equation}
R^{a}_{\,\,\, b\mu\nu}(\omega^a_{\ 
	b\mu})= \partial_\mu\omega^{a}_{\,\,\,b\nu}-
\partial_\nu\omega^{a}_{\,\,\,b\mu}
+\omega^{a}_{\,\,\,c\mu}\omega^{c}_{\,\,\,b\nu}-\omega^{a}_{\,\,\,c\nu}
\omega^{c}_{\,\,\,b\mu}\equiv 0.
\label{curvv}
\end{equation}
The most general connection satisfying this constraint is the pure gauge-like connection \cite{Obukhov:2002tm}
\begin{equation}
\omega^a_{\ b\mu}=\Lambda^a_{\ c} \partial_\mu \Lambda_b^{\ c},\label{connection}
\end{equation}
where $\Lambda_b^{\ c}=(\Lambda^{-1})^c_{\ b}$.

The torsion tensor of this connection 
\begin{equation}
T^a_{\ \mu\nu}(h^a_{\ \mu},\omega^a_{\ b\mu})=
\partial_\mu h^a_{\ \nu} -\partial_\nu h^a_{\ \mu}+\omega^a_{\ b\mu}h^b_{\ \nu}
-\omega^a_{\ b\nu}h^b_{\ \mu},
\label{tordef}
\end{equation}
is in general non-vanishing, and under the local Lorentz transformations
\begin{equation}
h'{}^a_{\ \mu}=\Lambda^a_{\ b}h^b_{\ \mu}, \quad \textrm{and} \quad
\omega'{}^a_{\ b\mu}=\Lambda^a_{\ c}\omega^c_{\ d\mu}\Lambda_b^{\ d}+\Lambda^a_{\ c} \partial_\mu \Lambda_b^{\ c},\label{lortrans}
\end{equation}
transforms covariantly. Therefore, this approach to teleparallel theories is called the covariant approach \cite{Krssak:2015rqa,Krssak:2015oua}; in contrast with the more common non-covariant approach  that will be discussed in section~\ref{weitz}.

The Lagrangian of teleparallel gravity is given by \cite{AP}
\begin{equation} 
\Lw_\text{TG}(h^a_{\ \mu},\omega^a_{\ b\mu})=\frac{h}{4 \kappa} T, \label{lagtot}
\end{equation}
where $h=\det h^a_{\ \mu}$, $\kappa=8\pi G$ is the gravitational constant (in $c=1$ units), and we have defined  the torsion scalar 
 \begin{equation}
 T= T(h^a_{\ \mu},\omega^a_{\ b\mu})= T^a_{\ \mu\nu}S_a^{\ \mu\nu},
 \label{tscalar}
 \end{equation}
where 
  $S_a^{\ \rho\sigma}$ is the superpotential 
 \begin{equation}
 S_a^{\ \mu\nu}=\frac{1}{2}
 \left(
 T^{\nu \mu}_{\ \ \ a}
 +T^{\ \mu\nu}_{a}
 -T^{\mu\nu}_{\ \ \ a}
 \right)
 -h_a^{\ \nu}T^{\sigma \mu}_{\ \ \sigma}
 +h_a^{\ \mu}T^{\sigma \nu}_{\ \ \sigma}.
 \label{sup}
 \end{equation}
A straightforward calculation shows that the teleparallel Lagrangian (\ref{lagtot}) and the Einstein-Hilbert one are equivalent up to a total derivative term \cite{AP}
\begin{equation}
\Lw_\text{TG} =\Lbol_\text{EH} -\partial_\mu \left(\frac{h}{\kappa} \, T^\mu \right), \label{lagequiv}
\end{equation}
from where follows that  the field equations of general relativity and teleparallel gravity are equivalent.

Since the Lagrangian (\ref{lagtot})  contains only the first derivatives of the tetrad, we can straightforwardly construct modified gravity theories with second order field equations. The most popular of these theories is the so-called  $f(T)$ gravity \cite{Ferraro:2006jd,Ferraro:2008ey,Bengochea:2008gz,Linder:2010py} defined by the  Lagrangian 
\begin{equation}
\Lw_f(h^a_{\ \mu},\omega^a_{\ b\mu})=\frac{h}{4\kappa}f(T),
\label{ftaction}
\end{equation}
where $f(T)$ is an arbitrary function of the torsion scalar (\ref{tscalar}).

The variation  with respect to the tetrad yields the field equations  \cite{Krssak:2015oua}
\begin{equation}
E_a^{\ \mu}(h^a_{\ \mu},\omega^a_{\ b\mu})=\Theta_a{}^{\mu}(h^a_{\ \mu},\Phi),\label{ftequation}
\end{equation}
where the term on the right-hand side is the energy-momentum tensor defined by
\begin{equation}
\Theta_a{}^{\mu} = \frac{1}{h} \frac{\delta \Lw_{\rm M}(h^a_{\ \mu},\Phi)}{\delta h^a{}_{\mu}} \,. 
\label{Theta}
\end{equation}
and $\Lw_{\rm M}(h^a_{\ \mu},\Phi)$ is the matter Lagrangian not depending  on derivatives of the tetrad. 
On the left-hand side we have introduced a shortened notation for the Euler-Lagrange expression
\begin{equation}\label{key}
E_a^{\ \mu}(h^a_{\ \mu},\omega^a_{\ b\mu})\equiv \frac{\kappa}{h}\frac{\delta \Lw_f(h^a_{\ \mu},\omega^a_{\ b\mu})}{\delta h^a_{\ \mu}},
\end{equation}
where
\begin{equation}
 E_a^{\ \mu}(h^a_{\ \mu},\omega^a_{\ b\mu})\equiv f_{TT}\, S_a^{\ \mu\nu} \partial_{\nu} T +
h^{-1} f_T \partial_{\nu}\left( h S_a^{\ \mu\nu} \right)
-f_T T^b_{\ \nu a }S_b^{\ \nu\mu}
+
f_T \omega^b_{\ a\nu}S_b^{\ \nu\mu}
+
\frac{1}{4}f(T) h_a^{\ \mu},
\end{equation}
and $f_T$ and $f_{TT}$ denote first and second order derivatives of $f(T)$ with respect to the torsion scalar $T$. The field equations of the ordinary teleparallel gravity are recovered in the case $f(T)=T$.

\section{Variational problem in teleparallel gravity \label{chvted}}
The difficulty of the variational problem in general relativity is  related to the presence of the second derivatives of the metric tensor (or tetrads in tetrad formalism) 
in the Einstein-Hilbert action. The straightforward variation of the action leads to the field equations that are satisfied only if we require vanishing  variations of both the metric and their normal derivatives on the  boundary. To satisfy both these requirements simultaneously is too strong condition that leads to the variation problem that is not well-posed, what is usually solved by adding the boundary term to the action  \cite{York:1972sj,Gibbons:1976ue,Dyer:2008hb}.

This problem is absent in teleparallel gravity as the teleparallel action contains just the first derivatives of the tetrad. However, we face a new difficulty since the 
action depends on the tetrad and the spin connection, and needs to be varied with respect to both.  

The straightforward variation of the teleparallel action  with respect to the spin connection as independent variable does not take into account the  teleparallel constraint (\ref{curvv}), and can be shown to be inconsistent \cite{Kopczynski1982,Krssak:2015lba,Golovnev:2017dox}. A viable solution  is to  enforce the teleparallel constraint (\ref{curvv}) through the Lagrange multiplier \cite{Kopczynski1975,Kopczynski1982,Blagojevic,Golovnev:2017dox}, but  this introduces  additional gauge symmetries associated with the multiplier  and the theory becomes significantly more complicated and hard to handle \cite{Blagojevic:2000pi,Blagojevic:2000qs}.

We follow here another approach that was introduced recently in Ref.~\cite{Krssak:2015lba}, which is based on writing the torsion scalar (\ref{tscalar}) in a form that  guarantees that the teleparallel constraint  (\ref{curvv}) is enforced. As it turns out, specifically in the case of the  torsion scalar (\ref{tscalar}) and the teleparallel connection  (\ref{connection}), we can write the torsion scalar as  \cite{Krssak:2015lba}
\begin{equation}	
T (h^a_{\ \mu},\omega^a_{\ b\mu})=
T (h^a_{\ \mu},0) + \frac{4}{h}B(h^a_{\ \mu},\omega^a_{\ b\mu}),  \label{rel}
\end{equation}
where we have introduced a shortened notation for
\begin{equation}
B=B(h^a_{\ \mu},\omega^a_{\ b\mu})=\,\partial_\mu \left(
h\, \omega^{a}_{\ b \nu}h_a^{\ \nu}h^{b\mu}\right)
\end{equation}.

We can then easily find that the variation  of the action with respect to the spin connection vanishes identically 
\begin{equation}
\frac{\delta \Lw_\text{TG}}{\delta \omega^a_{\ b\mu}}=\frac{\delta B}{\delta \omega^a_{\ b\mu}}=0,
\end{equation}
and   the field equations of teleparallel gravity (\ref{ftequation}) (with $f=T$) are in fact independent of the spin conneciton 
\begin{equation}
E_a^{\ \mu}(h^a_{\ \mu})=\Theta_a{}^{\mu}(h^a_{\ \mu},\Phi), \label{eqteg}
\end{equation} 
allowing us to solve the field equations using an arbitrary spin connection of the form (\ref{connection}).

Even though the spin connection does not affect the variations of the action, it  does   play an  important role in the theory. Beside ensuring the correct tensorial behavior under local Lorentz transformations (\ref{lortrans}), it is essential for  the correct definition of the conserved charges \cite{Obukhov:2006sk,Lucas:2009nq,Krssak:2015rqa}, and to obtain the physically relevant finite action \cite{Krssak:2015rqa,Krssak:2015lba}. 
To illustrate the latter one, let us  consider the teleparallel Lagrangian (\ref{lagtot}) for some  tetrad $h^a_{\ \mu}$ that solves the field equations and some arbitrary teleparallel spin connection of the form (\ref{connection}). We can  observe that  typically  the Lagrangian does not vanish at infinity and we obtain the  IR-divergent action \cite{Krssak:2015rqa}. 

The physical origin of this divergence can be understood   if we recall that the tetrad has 16 degrees of freedom and only  10  of them are related to gravity and the metric tensor (\ref{met}). The remaining  6  represent  the freedom to choose the frame, i.e.  the inertial effects,  which are not related to any actual physical fields and hence do not necessarily vanish at infinity. The torsion tensor that includes these spurious inertial effects does not vanish at infinity and leads  to the IR-divergent action.

The crucial observation is that  the teleparallel spin connection is related to the inertial effects as well (\ref{connection}), and it can be chosen  in a such way that it compensates the inertial effects associated with the tetrad. This results in the torsion tensor that does not include these spurious inertial effects and have correct asymptotic behavior
\cite{Krssak:2015rqa}.  To this end, we consider the   \textit{reference tetrad}, $\tref^{\;a}{}_{\mu}$, representing  the same inertial effects as the dynamical tetrad $h^a_{\ \mu}$. This  is achieved by ``switching off" gravity by setting some parameter that controls the strength of gravity to zero, e.g. in the case of the asymptotically Minkowski spacetimes, we can consider the limit
\begin{equation}
\tref^{\;a}{}_{\mu}\equiv \lim_{r \to \infty}   h^a_{\ \mu}, \label{reftet}
\end{equation} 
since  gravity as a physical field is expected to vanish at infinity. 

We then define the spin connection by the requirement that the torsion tensor vanishes for the reference tetrad 
\begin{equation}\label{torcond}
T^a_{\ \mu\nu} (\tref^{\;a}{}_{\mu},\omega^a_{\ b\mu})\equiv 0, 
\end{equation}
which has an unique solution
\begin{equation}
\omega^a_{\ b\mu} = \ombol^a_{\ b\mu}(\tref), \label{leveq}
\end{equation}
where $\ombol^a_{\ b\mu}(\tref)$ is the Levi-Civita connection for the reference tetrad. 

The torsion tensor $T^a_{\ \mu\nu} (h^{\;a}{}_{\mu},\omega^a_{\ b\mu})$ is then guaranteed to be well-behaved in the asymptotic limit by this construction, and the corresponding teleparallel action is  finite and  free of IR divergences  \cite{Krssak:2015rqa,Krssak:2015lba}. This  argument can be turned around, and  spin connection can be defined   by the requirement of the  finiteness of the action\footnote{This is in fact a  weaker requirement than (\ref{torcond}) since there exists a 1-parameter group of local Lorentz transformations that leaves the total derivative term in (\ref{rel}) invariant \cite{Ferraro:2014owa}.}. 

\section{Variational principle for modified teleparallel theories \label{chvft}} 
Let us now move to the modified case and consider the case of the covariant $f(T)$ gravity given by the Lagrangian (\ref{ftaction}), where we can use the relation (\ref{rel}) to straightforwardly vary the Lagrangian with respect to the spin connection
\begin{equation}\label{varspin2}
{\delta_\omega \Lw_f}=\frac{\partial B}{\partial \omega^a_{\ b\mu,\nu}} (\partial_\nu f_T) \delta \omega^a_{\ b\mu} =h\,h_a^{\ \mu}h^{b\nu}( \partial_\nu f_T) \delta \omega^a_{\ b\mu}.
\end{equation}
Using the  antisymmetricity of the spin connection we obtain
\begin{equation}\label{var11}
{\delta_\omega \Lw_f}=h h_{[a}{}^{\mu}h_{b]}{}^{\nu}(\partial_\nu f_T) \delta \omega^{ab}_{\ \ \mu}.
\end{equation}
For an infinitesimal local Lorentz transformation
\begin{equation}
\Lambda^a_{\ b}=\delta^a_{\ b}+\epsilon^a_{\ b}, \qquad \qquad \epsilon_{ab}=-\epsilon_{ba},
\end{equation}
the variation of the spin connection can be written as 
\begin{equation}
\delta \omega^{ab}{}_{\mu}=\D_\mu \epsilon^{ab}=\partial_\mu \epsilon^{ab} + \omega^a{}_{c\mu}\epsilon^{cb} + \omega^b{}_{c\mu} \epsilon^{ac}, 
\end{equation}
where $\D_\mu $ is the teleparallel covariant derivative acting on the algebraic indices only. We can then  integrate by parts to obtain 
\begin{equation}
{\delta_\omega \Lw_f}= (\partial_\nu f_T) \D_\mu\left(h h_{[a}{}^{\mu}h_{b]}{}^{\nu} \right)\delta \epsilon^{ab} \label{rel1}
\end{equation} 
which coincides with the result obtained recently using a different method of variation \cite{Golovnev:2017dox}. 

Assuming the ordinary matter with the Lagrangian independent of a spin connection, we can  set this variation to vanish for all $\epsilon^{ab}$ to obtain
\begin{equation}
Q_{ab}(h^{a}{}_{\mu},\omega^a_{\ b\mu})\equiv \partial_\nu T \,\D_\mu\left(h h_{[a}^{\ \mu}h_{b]}^{\ \nu} \right)=0, \label{concon}
\end{equation} 
where we have used $\partial_\nu f_T=f_{TT}\partial_\nu T$.

The relation (\ref{concon}) provides us  with 6 conditions relating the spin connection with the tetrad. We can observe that the spin connection enters this relation through the torsion scalar as well as through the teleparallel covariant derivative. We can check that the spin connection discussed in Ref.~\cite{Krssak:2015oua} for  the diagonal tetrads in spherical coordinate system in the case of  Minkowski or spherically symmetric spacetimes satisfy this condition.

Therefore, the situation in the modified case is radically different from the ordinary teleparallel gravity, where the spin connection is left undetermined from the field equations and can be fixed only by an  additional requirement of the finiteness of the action. In $f(T)$ gravity, the variation with respect to the spin connection is non-trivial and leads to the condition (\ref{concon}) for  the spin connection. It is interesting to observe that in many cases, e.g. diagonal tetrads for spherically symmetric spacetimes, the spin connection that satisfies the condition (\ref{concon}) also leads to the finite action. 

We would like to show that this is not a coincidence. In the case of the ordinary teleparallel gravity, the IR divergences appear only in the surface term that does not affect the variations of the action.  If we consider a Lagrangian to be a non-linear function of the torsion scalar, as we do in $f(T)$ gravity, divergences  appear not only in the surface term but  also in the ``bulk". This is a far more serious type of a divergence that leads to the failure of the variational principle and inability to derive field equations and the condition (\ref{concon}) consistently. We can now see that a finite action is a necessary condition for a consistent variational problem used to derive the  condition (\ref{concon}) itself.

\subsection{Case $T= \text{const}$ \label{seccons}}
An interesting situation is when the spin connection is chosen in a such way that $T= \text{const}$, which  straightforwardly  solves the condition (\ref{concon}). Naively, this seem to be in contradiction with the above statement about the finiteness of the action. However, it turns out that these solutions are in fact equivalent to the ordinary teleparallel gravity and hence avoid the problem of the divergences being in the ``bulk". 

The equivalence with teleparallel gravity can be seen if we restore the factor $f_{TT}$ in (\ref{concon}) and write
\begin{equation}\label{tconst}
f_{TT}\partial_\mu T= 0,   
\end{equation}
which we can rewrite then as
\begin{equation}
\partial_\nu f_T=0 \label{eq2}.
\end{equation}
The solution of this equation is a simple linear function
\begin{equation}
f=c_1 T +c_2,  \label{eq3}
\end{equation}
where $c_1$, $c_2$ are some constants, what return us back to the ordinary teleparallel gravity. 

We can that interpret this as that for any solution of general relativity it is possible to find such a spin connection that leads to $T= \text{const}$, which trivially  solves the field equations of  $f(T)$ gravity for an arbitrary function $f$ \cite{Tamanini:2012hg,Ferraro:2011ks,Bejarano:2014bca}. Therefore, solutions of general relativity can be considered to be universal solutions of arbitrary $f(T)$ gravity. However,  new solutions of $f(T)$ gravity can be obtained only if we find non-trivial solutions, i.e. not  $T= \text{const}$, to the condition  (\ref{concon}).

\section{Teleparallel theories as bigravity theories \label{bit}}
We have shown here that the  variation of the action with respect to both the tetrad and the spin connection lead us to the field equations (\ref{ftequation}) and the condition (\ref{concon}). 
However, solving both  simultaneously for a general tetrad and spin connection is usually a too complicated problem. 
We should remind here that the gravitational field equations are practically always solved using an  ansatz tetrad that corresponds to the symmetry of the problem. We now face an additional difficulty  of how to find the spin connection that solves the condition (\ref{concon}) for our ansatz tetrad. 

In Ref.~\cite{Krssak:2015oua}, it was suggested that the spin connection can be calculated by introducing the reference tetrad and using the relation (\ref{leveq}) to calculate the spin connection. This was motivated by the previous work in the ordinary teleparallel gravity \cite{Krssak:2015rqa}, where the reference tetrad--on the account of the property (\ref{eqteg})--could be defined self-consistently. In  $f(T)$ gravity it was argued that  the reference tetrad has to be ``guessed", what is obviously a rather ambiguous statement and needs to be elaborated.

We now present a consistent definition of the procedure developed in Ref.~\cite{Krssak:2015oua}. Since  the spin connection can be written as a function of the reference tetrad (\ref{leveq}), we can  straightforwardly re-write the field equations and the condition (\ref{concon}) as functions of two tetrads 
\begin{equation}
E_a^{\ \mu}(h^a_{\ \mu},\tref^{\;a}{}_{\mu})=\Theta_a{}^{\mu}(h^a_{\ \mu},\Phi), \qquad \qquad  Q_{ab}(h^a_{\ \mu}, \tref^{\;a}{}_{\mu})=0.  \label{elf}
\end{equation}
The reference tetrad  is  now determined from the system of equations (\ref{elf}) and the condition that the corresponding metric is the Minkowski one\footnote{We note  that we consider only the asymptotically Minkowski spacetimes here. The asymptotically (A)dS spacetimes require further analysis and will be addressed in the future.}
\begin{equation}
\gamma_{\mu\nu}=\eta_{ab}  \tref^{\;a}{}_{\mu} \tref^{\;b}{}_{\nu}. \label{metref}
\end{equation}
This results into a fully consistent definition of the theory, where the reference tetrad is determined self-consistently. There are two interesting aspects of why this reformulation in terms of two tetrads is worth considering. 

Firstly, working with two tetrads is often simpler for calculational reasons. As it was already demonstrated in Ref.~\cite{Krssak:2015oua}, it is relatively easy to  guess the reference tetrad from the form of the ansatz tetrad. We can then take this reference tetrad and perfrom a simply check whether the condition (\ref{concon}) is satisfied, what is a much simpler task than to solve it for a general teleparallel spin connection.

Secondly, it allows us to adopt a new perspective where  teleparallel theories  can be considered to be effectively bigravity theories. This is based on a simple observation that the presence of two tetrads effectively introduces two metrics to the theory;  the dynamical tetrad  defines the spacetime metric tensor $g_{\mu\nu}$ by  (\ref{met}) and  the reference tetrad  analogously yields  the background Minkowski metric (\ref{metref}).

This reveals some interesting insights into underlying degrees of freedom of these theories. We can recall that a general tetrad has 16 degrees of freedom, but the reference tetrad  is related to the Minkowski metric (\ref{metref}), which is uniquely determined by the choice of the coordinate system. Therefore, the metric degrees of freedom of the reference tetrad are fixed by the choice of coordinates and the condition of  $\gamma_{\mu\nu}$ being the metric of Minkowski space. The remaining 6  ``inertial" degrees of freedom are not fixed by the metric, and it is precisely these degrees of freedom--represented by the Lorentz matrix--that generate the teleparallel  spin connection through  the relation  (\ref{connection}).

We can also observe some interesting analogies and differences when compared with other well-known bimetric theories. Since the second metric is the   non-dynamical reference metric, there is some analogy with the  original bimetric theory of Rosen\footnote{The analogy with Rosen's theory is even more intriguing since a separation of gravity and inertia was  the original motivation of Rosen's bimetric theory, same as in the case of teleparallel gravity  \cite{Krssak:2015rqa,Krssak:2015lba}.  } \cite{Rosen:1940zza,Rosen:1940zz,Rosen:1974ua,Rosen:1980dp}, or  bimetric massive gravity \cite{Hassan:2011zd,Schmidt-May:2015vnx}.  However, there are also similarities with some recently proposed bigravity theories where the connection is determined from the second spacetime  \cite{Goenner:2010tr,Koivisto:2011vq,BeltranJimenez:2012sz}.

\section{Note on non-covariant teleparallel theories \label{weitz}}
The teleparallel connection (\ref{connection}) is a pure gauge-like, and hence there always exists a local Lorentz transformation, $\tilde{\Lambda}^a_{\ b}$, that transforms the connection to the vanishing one. In the original, non-covariant, approach to teleparallel theories \cite{Maluf:2013gaa,Ferraro:2006jd,Ferraro:2008ey,Bengochea:2008gz,Linder:2010py,Cai:2015emx}, the spin connection is gauged away independently of transforming the tetrad \cite{Krssak:2015oua}.

However, as our analysis in this paper shows, modified teleparallel theories are  consistent only if the spin connection satisfies condition (\ref{concon}) for the given tetrad.
Therefore, a local Lorentz transformation of the spin connection, should be always accompanied by a transformation of the corresponding tetrad 
\begin{equation}
\{h^a_{\ \mu},\omega^a_{\ b\mu}\} \xrightarrow{\hspace*{0.7cm}} \{\tilde{h}^a_{\ \mu},0\}, \label{wtransf2}
\end{equation}  
what reveals that only to a very special subclass of tetrads corresponds the vanishing zero connection. 

We would like to show that it is always possible to do the transformation (\ref{wtransf2}) and  formulate the theory directly  in terms of tetrads within this special class. We find that the tetrad must obey the constraint 
\begin{equation}\label{goodtet}
Q_{ab}(\tilde{h}^a_{\ \mu},0)=\partial_\nu T \,\partial_\mu\left(\tilde{h} \tilde{h}_{[a}^{\ \mu}\tilde{h}_{b]}^{\ \nu} \right)=0.
\end{equation}
This provides us with 6 conditions and  any tetrad that satisfies them is a ``good" tetrad in the sense of \cite{Tamanini:2012hg}. For instance, one can check that the off-diagonal tetrad for a spherically symmetric problem given by Eq. (39) in Ref.~\cite{Krssak:2015oua} satisfy this condition. As far as we are interested in the solutions of the field equations only, we obtain the very  same results as in the covariant approach. 

The  non-covariant teleparallel theories can be viewed then as teleparallel theories formulated in a  particular gauge, analogously to  electromagnetism formulated in some particular gauge. The loss of local Lorentz invariance is then not a sign of any pathology, and is indeed analogous to loosing gauge symmetry after fixing the gauge.  More serious drawback exposed in this paper is that the condition for a good tetrad (\ref{goodtet}) cannot be derived naturally in the non-covariant framework, while in the covariant approach the condition (\ref{concon}) is derived straightforwardly by varying the action with respect to the spin connection.

\section{Conclusions}
We have analyzed the variational principle in the covariant formulation of teleparallel theories, where the fundamental variables are the tetrad and the  spin connection subjected to the teleparallel constraint (\ref{curvv}). In the ordinary teleparallel gravity the spin connection does not affect the field equations due to the fact that it enters the action through the surface terms only.  The spin connection can be then fixed just by some additional  requirements as is the finiteness of the action or condition of obtaining the physically relevant  conserved charges \cite{Krssak:2015rqa}. 

We have shown that in the modified case the situation is radically different and the variation of the action with respect to the spin connection is non-trivial (\ref{rel1}); coinciding with the recent result obtained using a different method of variation in Ref.~\cite{Golovnev:2017dox}.  We have then demonstrated that for the ordinary matter, this non-trivial variation provides us with condition (\ref{concon}) that completely determines the spin connection and relates it with the tetrad. We have argued  that the solutions of this condition lead to the finite action, what is a necessary condition for a consistent variational principle.  As a special case, we have shown that in the case $T=\text{const}$ the $f(T)$ gravity reduces to the ordinary general relativity what explains the universality of its solutions \cite{Ferraro:2011ks,Bejarano:2014bca}. 

We have  demonstrated that in the usual (non-covariant) formulation, the condition (\ref{concon}) can be shown to be equivalent to a new condition for  the tetrad (\ref{goodtet}). Any tetrad that solves this condition is guaranteed to be ``good" in the sense of Ref.~\cite{Tamanini:2012hg}, what solves the long-standing problem of how to choose the approapriate tetrad in the non-covariant formulation of $f(T)$ gravity.

We have then argued that it is possible to re-formulate teleparallel theories  in terms of the so-called reference tetrad instead of the spin connection.  If we adopt this viewpoint, modified teleparallel theories can considered to be effectively bigravity theories,  where the dynamical tetrad generates the spacetime metric, while the reference tetrad introduces the background reference metric and its inertial components generate the spin connection. This is a novel viewpoint on modified teleparallel theories and opens new way to study their underlying physical structure, as well as intriguing analogies with other bimetric theories of gravity.

Our results were demonstrated on the example of $f(T)$ gravity, where we could perform the variation with respect to the spin connection relatively easily due to the relation (\ref{rel}). To extend  these results to other modified teleparallel  theories  with second order field equations, e.g. see \cite{Hayashi:1979qx,Maluf:2011kf,Bahamonde:2017wwk}, we need another  method to vary the action with respect to the  spin connection subjected to the teleparallel constraint (\ref{curvv}). We expect that the constrained Palatini formalism  should  be applicable in the general case and should lead to a condition on the spin connection analogous to Eq.~(\ref{concon}). Another approach is to include derivatives of torsion  in the action. At least in  certain cases it is possible to show that  the spin connection does not need to be restricted by any constraint \cite{Bahamonde:2015zma}, but generally  this leads to the fourth order field equations what introduces  difficulties known from the curvature-based modified theories.

\section{Acknowledgments}
The author would like to thank S.~Bahamonde, R.~Ferraro, A.~Golovnev, M.-J. Guzm\'an, M.~Hohmann, L.~J\"arv, K.~Koivisto, L.-W.~Luo, J.~N.~Nester, J.~W.~Maluf, J.~G.~Pereira, C.~Pfeifer, E.~N.~Saridakis,  H.-H.~Tseng, Y.-P.~Wu for interesting and stimulating  discussions, and  C.-Q.~Geng for hospitality during the stay at NCTS (Hsinchu, Taiwan).  This research is funded by the European Regional Development Fund through the Centre of Excellence TK 133 \textit{The Dark Side of the Universe}.

\bibliography{TeleBib}
\bibliographystyle{Style}

\end{document}